\documentclass[a4paper, onecolumn, amsfonts, amssymb, amsmath, preprint, showkeys, nofootinbib, superscriptaddress,twoside]{revtex4-2}
\usepackage[english]{babel}
\usepackage[utf8]{inputenc}
\usepackage[colorinlistoftodos, color=green!40, prependcaption]{todonotes}
\usepackage{setspace} 
\usepackage{upgreek}
\usepackage{comment}
\usepackage[pdftex, pdftitle={Article}, pdfauthor={Author},hidelinks]{hyperref}
\usepackage{setspace}
\usepackage{lineno}

\def\be{\begin{equation}}
\def\ee{\end{equation}}
\def\e#1{\label{#1}\end{equation}}
\def\bea{\begin{eqnarray}}
\def\eea{\end{eqnarray}}
\def\ea#1{\label{#1}\end{eqnarray}}

\def\bem#1{\begin{mathletters}\label{#1}}
\def\eml{\end{mathletters}}

\def\ket#1{{|#1\rangle}}
\def\bra#1{{\langle#1|}}
\def\ketbra#1{{|#1\rangle}{\langle#1|}}

\def\mean#1{{\langle#1\rangle}}
\def\4#1{{\boldsymbol{#1}}}
\def\8#1{{\widetilde{#1}}}
\def\bse{\begin{subequations}}
\def\ese{\end{subequations}}

\begin{document}
\title{Anti-Zeno purification  of spin baths by quantum probe measurements} 
\author{Durga Bhaktavatsala Rao Dasari} 
\email[Corresponding author:]{d.dasari@pi3.uni-stuttgart.de}
\affiliation{3.\,Physics Institute, University of Stuttgart, Center for Applied Quantum Technologies,
IQST, MPI for Solid State Research, Stuttgart 70569, Germany.}
\author{Sen Yang} 
\affiliation{Department of Physics, The Hong Kong University of Science and Technology, Clear Water Bay, Hong Kong, China.}
\author{Arnab Chakrabarti}
\affiliation{AMOS and Department of Chemical and Biological Physics, Weizmann Institute of Science, Rehovot, Israel}
\author{Amit~Finkler}
\affiliation{AMOS and Department of Chemical and Biological Physics, Weizmann Institute of Science, Rehovot, Israel}
\author{Gershon Kurizki}
\affiliation{AMOS and Department of Chemical and Biological Physics, Weizmann Institute of Science, Rehovot, Israel}
\author{J\"org Wrachtrup}
\affiliation{3.\,Physics Institute, University of Stuttgart, Center for Applied Quantum Technologies,
IQST, MPI for Solid State Research, Stuttgart 70569, Germany.}

\begin{abstract} 
 \noindent The quantum Zeno and anti-Zeno paradigms have thus far addressed the evolution control of a quantum system coupled to an immutable bath via non-selective measurements performed at appropriate intervals. We fundamentally modify these paradigms by introducing, theoretically and experimentally, the concept of controlling the bath state via selective measurements of the system (a qubit). We show that at intervals corresponding to the anti-Zeno regime of the system-bath exchange, a sequence of measurements has strongly correlated outcomes.  These correlations can dramatically enhance the bath-state purity and yield a low-entropy steady state of the bath. The purified bath state persists long after the measurements are completed. \textcolor{black}{Such purification enables the exploitation of spin baths  as long-lived quantum memories  or as quantum-enhanced sensors.  The experiment involved a repeatedly  probed defect center dephased by a nuclear spin bath in a diamond at low-temperature.}

\end{abstract}


\maketitle
\newpage
\section*{Introduction}

Quantum system interactions with their environment, alias a ``bath" that has many degrees of freedom, are foundationally interesting since they chart the path from quantum coherent dynamics to behavior described by statistical-mechanics or thermodynamics as the complexity of the system-bath compound increases \cite{ref2, ref2.5, ref3, ref5, ref8}. On the applied side, adequate control over system-bath interactions is a prerequisite for all emerging quantum technologies \cite{ref10,ref11,ref13,ref14,ref15,ref16,ref17,ref18,ref19,ref20,ref21,ref22,ref23,ref24,ref24.10,ref24.50,ref24.55,ref25,ref26,ref27,ref28,ref29,ref32,ref33}. In particular, fault-tolerant quantum computation \cite{ref11} requires the ability to protect the information encoded in the computational qubits from leaking into the environment (bath), thus causing their decoherence. The desirable, often unachievable, goal is to completely decouple the system from the bath \cite{ref35,ref37,ref38,ref39,ref39a,ref39b}. More realistically, the adverse effects of decoherence can be suppressed by control pulses at a rate faster than the inverse memory (correlation) time of the bath \cite{ref40,ref41,ref42,ref44,ref46}. Yet, existing system-bath interaction control is severely constrained by the ruling paradigm whereby the bath is immutable and control is only applicable to the system \cite{ref2,ref2.5,ref35,ref37,ref38,ref39,ref39a,ref39b,ref40,ref41,ref42,ref44,ref46}. The justification	is that in typical experimental settings one can only achieve a high degree of control and readout precision of the system, but not of \textcolor{black}{a spectrally unresolved, randomly fluctuating bath. In contrast, a high degree of control has been achieved \cite{ref22, ref27} in the preparation and readout of spectrally resolvable baths consisting of few nuclear spins in a diamond, by means of probe-induced control gates among the bath spins and their conditional post-selected \cite{ref22, ref27} initialization -- methods that are difficult to extend to large, spectrally unresolvable random baths.}

\textcolor{black}{Here we capitalize on the major stride, that has been made since the year 2000,} towards deeper understanding of quantum aspects of system-bath interactions: the discovery that the quantum Zeno effect (QZE) \cite{ref48} and its inverse, the anti-Zeno effect (AZE), can profoundly affect open quantum systems \cite{ref50,ref51,ref52,ref53,ref54}. These effects occur on time-scales such that system-bath interaction dynamics may allow for partial revival of the system coherence caused by energy and entropy backflow from the bath to the system. The most familiar manifestations of the QZE or the AZE are, respectively, the slowdown or speedup of the open- system state- evolution when subjected to measurements at appropriate rates \cite{ref46,ref50,ref51,ref52, ref53, ref54, ref55,ref56,ref57,ref58,ref59}. This evolution control is effected by non-selective measurements  of the system (i.e., disregarding their outcomes).

Not less significant has been the discovery that frequent non-selective measurements (NSM) in the  QZE regime can heat up a qubit, whereas less frequent NSM conforming to the AZE regime can cool the qubit down  while  the bath remains unchanged \cite{ref60,ref61}. Such  measurement-induced cooling and heating effects have been experimentally verified by some of us for interacting spin-systems \cite{ref62}.

Here we fundamentally modify the QZE-AZE paradigms by: (i) rendering the bath controllable by measuring the system and (ii) employing selective instead of non-selective measurements of the system. Concretely, we introduce, theoretically and experimentally, the concept of controlling the bath state via selective measurements of the system (a qubit) at intervals corresponding to the AZE regime of the system-bath exchange. We show that a sequence of measurements at intervals that correspond to the AZE regime renders the bath evolution conditional on the strongly correlated outcomes of this sequence of measurements. Such a sequence can be viewed as a conditional trajectory (CT) in parameter space. Upon choosing a particular CT, the bath is forced to be purified to a desired low-entropy steady-state which persists long after the measurements have ceased. The present protocol is a conceptual addition to the arsenal of measurement- induced quantum state engineering that has been theoretically proposed to generate exotic entangled states of spin ensembles \cite{ref63} as well as cooling or squeezing of cantilever modes \cite{ref64}. Those schemes, in contrast to the present one, have not relied on the dependence of state preparation on measurement intervals, which play a crucial role here.

We consider  the largely unexplored domain of strong system-bath coupling for which a selective (projective) measurement of the system can drastically disturb the bath state depending on the measurement outcome and the interval between measurements. This disturbance becomes stronger as the bath gets smaller, but is strong enough to allow the control of baths composed of $N \gg 1$ spins, as we have verified. Existing control methods that rely on the weak-coupling (Born, even if not Markov) approximation \cite{ref33,ref35, ref37, ref38, ref39, ref39a, ref39b, ref40, ref41, ref42, ref44, ref46} are inadequate in such situations. Instead, one has to rely on the exact solution of the Hamiltonian evolution. Yet, exact solutions are obtainable for only a few system-bath interaction models.

We here theoretically and experimentally investigate such an exactly solvable model: the dephasing of a two-level system (probe spin) by a degenerate (non-interacting) spin bath with conserved total spin. The protocol consists in the preparation of the probe spin in a superposition of its energy states, followed by frequent observation of the decay (dephasing) of this superposition via the probe interaction with an adjacent, initially thermal, spin bath. The idea \textcolor{black}{(Fig.\,1a)} is to repeatedly check via frequent selective measurements whether the probe is still in its initial state (a positive outcome, corresponding to $0$ photon emission by the probe) or has decayed already (a negative outcome, corresponding to $1$ photon emission).  The novelty here is that depending on the interval between measurements, only certain bath states contribute to the preservation of the initial probe state. This dependence on the interval between measurements with selected outcomes brings about gradual purification	of the mixed bath state, since those measurement outcomes are incompatible with most of the bath eigenstates in the mixture. After a sequence of few measurements with selected outcomes at appropriate intervals (a chosen CT), the bath reaches a  low-entropy steady state. By selecting the desired sequence we may control the purified steady state of the bath and thereby, long after the purification control has ceased, imprint on demand, the desired evolution on qubits immersed in this bath: either dramatically prolonged coherence time, corresponding to the QZE regime, or, conversely, rapid oscillations of the  qubit coherence, in accordance with the AZE regime.

As opposed to the existing NSM-based paradigm, whereby the transition from the QZE to the AZE regime occurs by extending the interval between measurements, here the QZE-AZE transition can also be effected by switching from a CT with only positive outcomes to a CT with both negative and positive outcomes, since these \textcolor{black}{CTs} correspond to very different  steady states of the bath (cf.\,Results, Supplementary Note 3). The surprising advantageous feature of the considered measurement sequences is that their success probability stops decreasing with the number of trials, as opposed to the KLM protocol \cite{klm} \textcolor{black}{(Fig.\,1b)}. \textcolor{black}{The autocorrelation time, over which there is almost perfect correlation between the results of measuring the coherence of the qubit at different times, is increased dramatically after the bath purification: it is shown in our experiment to be typically $\sim 10^3$ times longer in a purified bath, than in an unpurified bath. We stress that our measurements of the probe are strong, i.e. projective, as opposed to the previously employed weak measurements \cite{ref24,ref24.10,ref24.50,ref24.55}.}

\begin{figure}[!h]
\begin{center}
\includegraphics[width=0.9\textwidth]{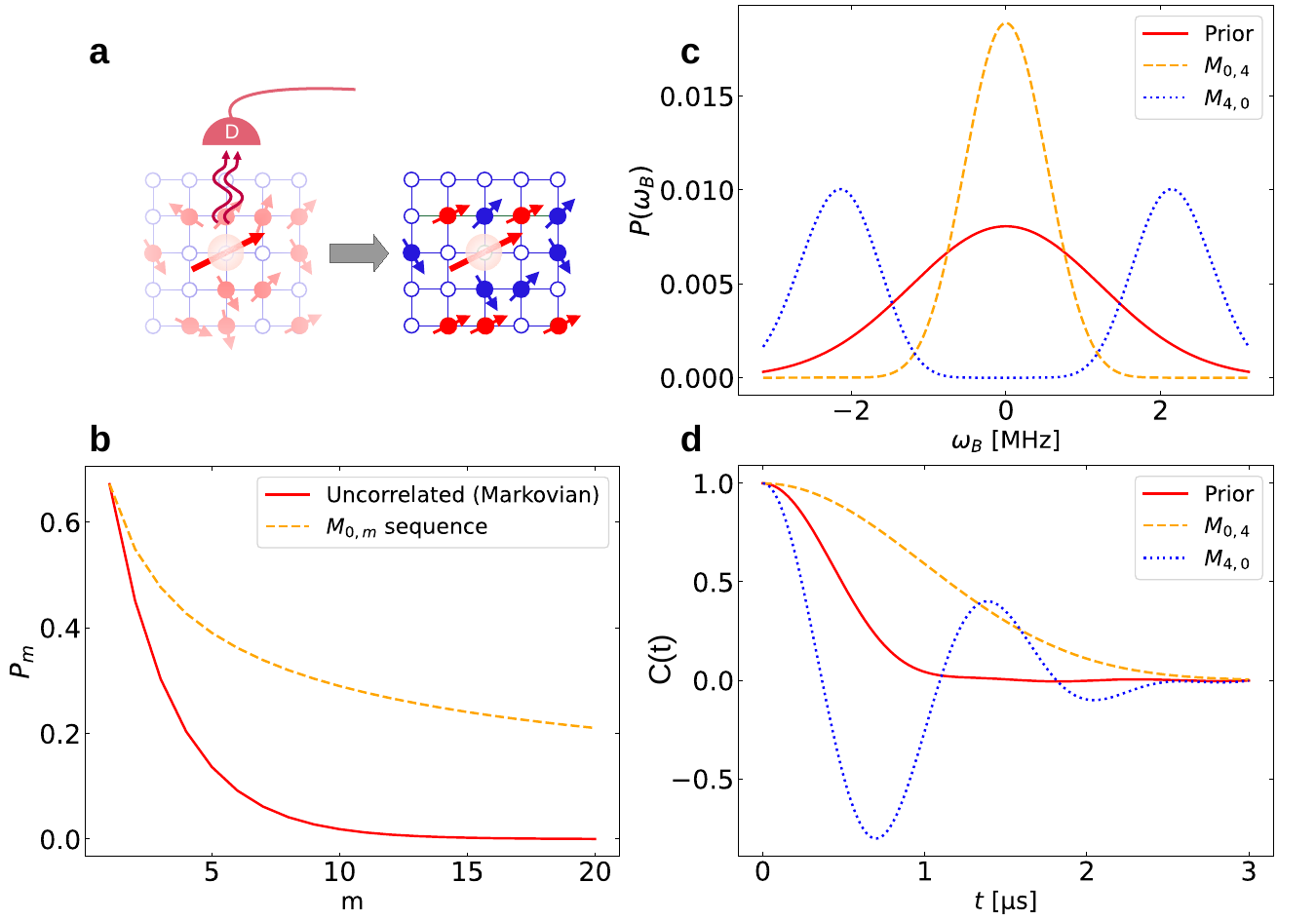}
\end{center}

\vspace{-5mm}
\caption{{\bf Bath purification by conditional selective measurements and its posterior (persistent) effects} (a) Schematic description: Conditional selective measurements of the state of the central probe-spin (S) \textcolor{black}{surrounded by a spin bath in the star configuration. The conditional measurements consist of photon-emission detection from the decayed state or non-detection from the initial state. A sequence of measurement events} collapses the spin-bath toward a low entropy state with resolvable, partly-\textcolor{black}{(red)} or \textcolor{black}{fully-polarized} (orange) spins. (b) The success probability, $p_m$ of $m$ positive measurement outcomes: solid red -- for a probe subject to uncorrelated (Markovian) measurement sequence (solid red) that leave the spin-bath unchanged, dashed orange -- for a probe subject to  correlated measurements that purify the bath. The success probability exponentially decays with $m$ for uncorrelated measurements, while it saturates for correlated measurements. (c) Simulated bath spectrum $P(\omega_j)$ plotted for the cases prior to measurements (red -- thermal bath) and after the conditional (selective) measurement sequences (CT) $M_{0,4}$ (orange) and $M_{4,0}$ (blue).  Both (orange and blue) posterior spectra are much narrower than the prior spectrum, indicating low-entropy steady-states of the bath. The orange spectrum is peaked at bath states with zero magnetic field, whereas the blue spectrum is peaked at higher magnetic field values of the bath spins. (d) Probe coherence decay  prior (red) and posterior to measurements (orange, blue) for the spectra shown in (c). The orange curve is in the quantum Zeno (QZE) regime (decay slowdown) while the blue curve is in the anti-Zeno (AZE) regime (decay speedup). These curves are simulated for a bath of $N=8$ spins that are randomly (inhomogeneously) coupled to the probe spin. The timescale between measurements is chosen to be $\tau = 0.6\,\upmu$s, such that it is close to the bare $T_2^*$ obtained from the prior FID (red solid-line in 1(d)). (See Results, Supplementary Note 3)}
\label{BathPure}
\end{figure}

\section*{Results}

\noindent
\subsection*{Theory}

We consider the dipolar interaction of an (effectively) spin-$1/2$ probe system ($S$) with $N$ identical spin-$1/2$ systems (the bath – $B$). The inherent quantization axis of $S$ and the $S$-$B$ resonance mismatch (detuning) render their interaction dispersive, which leads to decoherence (pure dephasing) of the $S$-spin. Commonly, this problem involves spin-spin interactions in the bath, as in spin chains \cite{ref44, ref75, ref76}.  The dynamics generated by all such Hamiltonians can be exactly solved for any distribution of $S$-$B$ couplings and any eigenvalue spectrum of $B$. Here we assume, consistently with our experiment, \textcolor{black}{that all the bath $B$ spins are directly coupled to the probe-spin $S$, while inter-spin interactions within the bath are absent. This condition, realized in our experimental setup, is commonly termed as the ``star configuration" (Fig.\,\ref{BathPure}a).} The Hamiltonian describing the $S$-$B$ interaction is then given by \textcolor{black}{(cf.\, Supplementary Note 1)}

{\color{black}{}

\bea
H &=& \hat{S}_z \otimes \hat{B},\\ \nonumber
\hat{B}&=& \sum_k {\rm{\bf{g_k}}} \cdot {\rm{\bf{I_k}}}.
\label{eq:basic_H}
\eea

Here $\hat{S}_z$} is the $S$-spin \textcolor{black}{Pauli-operator} component along its quantization axis, $k$ labels the $B$-spins, ${\rm{\bf{g_k}}}$ are the dipolar couplings  between $S$ and the $k$-th nuclear spin, and ${\rm{\bf{I_k}}}$ its spin (Pauli) operator. The typically random spatial locations of the $B$-spins result in an inhomogeneous  distribution of the couplings ${\rm{\bf{g_k}}}$ such that $B$ has no preferred spatial symmetries. 

In the $\hat{S}^z$ basis of $S$ spanned by states ($|{e}\rangle, |{g}\rangle$), the above Hamiltonian can be rewritten as 
\bea
H= \hat{B}_e \ket{e}\bra{e} +\hat{B}_g\ket{g}\bra{g}, ~~ \hat{B}_{e(g)} = \pm \sum_k {\rm{\bf{g_k}}}(r) \cdot {\rm{\bf{I_k}}}
\label{Eq:H_in_Sz_basis}
\eea 
where $\hat{B}_{e(g)}$ are the spin-bath  energy operators associated with the energy states $\ket{e(g)}$ in $S$. 
The  Hamiltonian in Eq.\,\ref{Eq:H_in_Sz_basis} gives rise in the interaction picture to the closed-form time-evolution operator of the ``supersystem'' $S+B$  
\be
\label{uop}
\hat{U}_{S+B}(t) = \hat{U}_e(t)|e\rangle\langle e| \, + \, \hat{U}_g(t)|{g}\rangle\langle g|
\ee 
where $\hat{U}_{e(g)}(t) = {\exp{(it \hat{B}_{e(g)})}}$ govern the dynamics of $B$ in a manner conditional  on the state of $S$.  In what follows, we review standard dephasing that assumes an unchanging statistical state of $B$ and contrast it with the conditional dynamics of $B$ under correlated measurements.


In standard treatments, the  effect of $B$ on $S$ is mimicked by a classical, noisy (random) magnetic field, assuming the initial $B$-state to be completely depolarized (mixed): the $S$-$B$ interaction given above does not lead, then, to any bath dynamics, i.e.,
  \be
  \rho_B(t) = \rho_B(0), ~~\rho_B(0) = \sum_j P_j(0) \ketbra{j}, P_j(0) = \mathcal{N}\exp{(-\omega_j^2/2N)}
  \ee
where $j$ labels the collective $B$-eigenstates that satisfy the eigenvalue equation  $B\vert j\rangle = \omega_j\vert j\rangle$ and $\mathcal{N}$ is the appropriate normalization constant. If the eigenvalues $\omega_j$ form a continuum, then $\mathcal{N} = \sqrt{\frac{1}{2\pi N}}$. For discrete spectra, $\mathcal{N}$ has to be calculated by direct summation.  

As the  $S$-$B$ interaction causes pure dephasing (decoherence) of $S$, the dynamics only occurs between superposition states of $S$ in the $e(g)$ basis, say, $\ket{\pm} = \frac{1}{\sqrt{2}}[\ket{e} \pm \ket{g}]$. 
If we initialize the probe in the $\ket{+}$ state, we can measure its free-induction decay (FID) \cite{ref39}, caused by its mixing with the orthogonal superposition state $\ket{-}$ via a bath-induced random phase. The ensemble average  of their populations corresponds to the probabilities of measuring  these states at time $t$, given by $P_+ (t)  \equiv 1-P_-(t) = \frac{1}{2}[1+{\rm e}^{-(t/T_2)^2}]$. \textcolor{black}{The decoherence rate $1/T_2$ is the root mean square (rms) of the probe-bath coupling $\bar{g}$. If the spin bath is the only noise source, it is the inverse bare decoherence time.} 
Equivalently, the probe-spin coherence can be measured by the decaying mean value of the $\mean{S^x}$ component of its spin, which displays Gaussian decay, given by
$\mean{\hat{S}^x}(t) \approx {\rm e^{-(t/T_2)^2}}$. 

\textcolor{black}{The dephasing (decoherence) time $T_2$ can be stretched by dynamical control of $S$ aimed at minimizing the noise effects. A common (but not always optimal) example of such control is dynamical - decoupling (DD) that uses pulse sequences \cite{ref35,ref38,ref39,ref37,ref66} to shift the spectral response of the system beyond that of the bath, but is blind to the bath spectrum \cite{ref35,ref37,ref38,ref39}. An alternative is the bath-optimal minimal-energy control (BOMEC) strategy that employs quasi-continuous control to anti-correlate the system and the bath spectra, thereby taking advantage of the gaps and dips encountered in diverse bath spectra \cite{ref42}.}

Here we set out to overcome the \textcolor{black}{inherent} limitations of any form of conventional dynamical control of $S$, be it DD or BOMEC: (i) the high rate of pulses required to effectively control decoherence in the high- temperature or broad-band (white noise) limit of $B$, which are properties that cannot be affected by conventional \textcolor{black}{dynamical} control; (ii) the need to maintain the dynamical control during the time that other\textcolor{black}{, quantum-logic or quantum-sensing operations,} are performed on the qubits of interest (although this difficulty can be mitigated by appropriate control \cite{ref67}). To overcome these limitations, we resort to unconventional bath control via conditional selective measurements of the probe.  

Let $\tau$ be the interval between consecutive measurements of $S$. If the interval greatly exceeds the bath correlation (memory) time, $t_c \ll \tau$, then the regime is Markovian, i.e., the results of each measurement of $S$ are independent of its predecessors. This means that $S$ interacts every time with $B$ that is in the same statistical (thermal) state. Then, following many repetitions of the FID measurement, one can extract the average coherence at a specific time to infer the value of the dephasing time $T_2$.
We, in contrast, wish to venture into the opposite, strongly non-Markovian limit, that corresponds to highly correlated results of consecutive measurements. Namely, the readout at the time $t = m\tau$ ($m$ is an integer) depends on the outcomes of all its predecessors. In this limit, the $B$-state evolution is governed by the non-unitary evolution operator corresponding to $S$ projections onto either of the $\ket{\pm}$ states, given by {\color{black}(cf.\, Supplementary Note 1)}

\be 
\hat{V}_\pm(t) = (\hat{U}_e (t)\pm \hat{U}_g(t))/2, ~~\hat{U}_{e(g)} = {\rm e}^{\pm i\omega_j t}\ketbra{j}.
\label{eq:Vpm}
\ee
The conditional evolution of $B$ assumes the form 
\be
\rho^\pm_B(t) \equiv \hat{V}_{\pm}(t)\rho_B(0)\hat{V}^\dagger_{\pm}(t)=\sum_j P_j(t)\ketbra{j},
\label{eq:rhoBpm}
\ee
where $\ket{j}$ are the collective eigenstates of the bath operator in Eq.\,\ref{Eq:H_in_Sz_basis}, and $P_j(t)$ are conditionally modified populations {\color{black}(cf.\, Supplementary Note 2)}. Repeated application of the conditional evolution operator in Eqs.\,\ref{eq:Vpm}-\ref{eq:rhoBpm} can yield a purified state of $B$ corresponding to the bath spectrum $G = \sum_j \tilde{P}_j(\omega_j)$, where $\tilde{P}_j$ is the steady-state probability of the $j^{\rm th}$ eigenstate with frequency $\omega_j$.

Let us consider that $m$-fold measurements of $S$ are consecutively performed in the $\ket{+}, \ket{-}$ basis. There are $2^m$ measurement sequences with different possible outcomes. The projections on the $\ket{+}$ or $\ket{-}$ state in each of these (single-shot) measurements are designated as $'0'$, $'1'$ respectively. Each such sequence, denoted by $M_{n,m}$,  can be represented as an $n$-fold string of ‘1’ and $m$-fold string of ‘0’. The bath evolution is conditional on a particular sequence or string, which, in parameter space, can be viewed as a trajectory – dubbed henceforth ``conditional trajectory'' (CT).  

For a CT associated to the sequence $M_{n,m}$, performed at intervals $\tau$, the corresponding non-unitary $B$-evolution operator is (see Supplementary Note 1,2)

\be\label{V1}
\hat{V}_{n,m}(t = m \tau) = \cos^{m-n}(\hat{B}t)\sin^n (\hat{B}t).
\ee 
The  $j$-distribution that underlies the $B$-state is progressively  narrowed down  compared  to that of the initial thermal state, Eq.\,\ref{eq:basic_H}, in a manner that depends on the realized CT. This comes about since, at a chosen time $t$, different $j$-states will have either maximized or minimized probabilities, depending on the arguments of the respective cosine or sine functions. The outcomes are correlated provided $\tau \ll T_2$, resulting in the bath state

\be\label{V2}
\rho_B(t = m\tau, M_{n,m}) = \frac{\hat{V}_{n,m}(t)\rho_B(0)\hat{V}_{n,m}^\dagger(t)}{{\rm Tr}[\hat{V}_{n,m}(t)\rho_B(0)\hat{V}_{n,m}^\dagger(t)]}. 
\ee 
\textcolor{black}{An important consideration is the success probability: the fraction of measurement sequences that yield the desired outcomes described by $M_{n,m}$.}
In particular, for the CT $M_{0,m} = \lbrace{0,0,\cdots 0 \rbrace}$ at intervals $\tau$, the CT success probability $p_{0,m} (t = m\tau) ={{\rm Tr}[V_{0,m}(t)\rho_B(0)V_{0,m}^\dagger(t)]}$ is maximized for $\omega_j\tau \simeq l\pi$ ($l$ integer), because of the $\tau$-dependence of $\hat{V}_{0,m} = \cos^m(\hat{B}\tau)$. Yet any correlated CT in Eq.\,\ref{V1} with appropriate intervals $\tau$ has the remarkable property (Fig.\,\ref{BathPure}b) that its success probability saturates (stops decreasing) with the number of measurements, $m$, unlike the exponentially decreasing probability of an uncorrelated CT sequence, as in the KLM protocol \cite{klm}. By selecting a CT in Eq.\,\ref{V1},\ref{V2} a specific steady state of $B$ is reached with appreciable success probability. These steady states are given by $\rho_B(M_{n,m}, t\rightarrow \infty) = \frac{1}{2^{N^\prime}}\sum_{j^\prime=-N^\prime/2}^{N^\prime/2}\ketbra{{j^\prime}}$,
where $\ket{j^\prime}$ are the $B$-eigenstates that have ``survived'' the purification by the selected CT and ${N^\prime} \ll N$ and therefore can have much higher purity than its initial high-temperature state.

It is appropriate \textcolor{black}{(cf.\,Supplementary Note 3, 4)} to choose for the purification the time-interval of resolvable measurements that conforms to AZE regime, \textcolor{black}{$1/\bar{g} \geq \tau > 0.1/\bar{g}$, where $\bar{g} = \sqrt{\sum_k g_k^2}$ is the rms S-B coupling}. Then the steady state of $B$ following purification according to the $n = 0$ CT in Eq.\,\ref{V2} tends to a single-peak distribution around the zero-field $\vert j = 0\rangle$ state. By contrast, the CT with $n = 4$ in Eqs.\,\ref{V1},\ref{V2} yields a statistical mixture of total-magnetization (macroscopic) states that are maximally separated in energy,  

\be
\rho_B(t\rightarrow \infty) =\frac{1}{2}[\ketbra{j=-\frac{N}{2}} + \ketbra{j=+\frac{N}{2}}].
\label{eq:rho_infty}
\ee
These steady states dramatically raise the bath purity that starts from $1/2^N$. The individual spins remain unpolarized, since the steady-state polarization is not local, but resides in the collective spin-angular momentum states. 

The steady-state of $B$ following such purification is a persistent imprint left by the corresponding CT. This imprint is revealed by initializing a probe qubit in the $\vert + \rangle$ state and tracking the subsequent evolution of its mean coherence by repeated measurements with selected time-intervals. The present protocol allows a different perspective of the Kofman-Kurizki (KK) universal formula \cite{ref40,ref46,ref50,ref51} for dynamically controlling the decoherence of $S$ that is weakly coupled to $B$, whereby the $S$-coherence obeys:

\be
 \mean{\hat{S}^x}(t) \sim \exp{\Big[-\int G(\omega)F(t,\omega)d\omega\Big]}
\label{Eq:KK_formula}
\ee
$F( t,\omega )$ being the time-dependent filter-function shaped by the power-spectrum, which may arise from unitary (pulsed or continuous) operations or from frequent measurements, and $G (\omega)$ is bath-coupling spectrum. \textcolor{black}{In the present study, $F(t,\omega)$ depends on the choice of the CT, $M_{n,m}$. Since the measurement interval is here fixed, the $F(t,\omega)$ spectral filter is unaffected by permutations of the $n$ and $m$ outcomes (cf.\,Supplementary Note 2).} 

For the initially thermal spin-bath, $G(\omega) \propto \exp(-\omega^2/\Delta_T^2)$, where $\Delta_T$ is the temperature-dependent spectral width. According to the KK formula \textcolor{black}{\cite{ref2.5, ref40, ref41, ref42, ref44, ref46}}, the goal of existing dynamical control is to adapt the $F(t,\omega )$ filter spectrum to the immutable $G (\omega)$ spectrum so as to either suppress or enhance the decoherence in Eq.\,\ref{Eq:KK_formula}, corresponding to the QZE or AZE regime, respectively. Bath-optimized minimal energy control (BOMEC) \cite{ref42} follows the same procedure, but can be much more effective, since it yields an optimal multi-dimensional filter that can adapted to a probe coupled to different baths simultaneously. Here, in contrast to all existing control methods, the KK formula can be employed differently: a chosen $M_{n,m}$ sequence can dramatically modify the filter spectrum $F(t,\omega)$ in Eq.\,\ref{Eq:KK_formula} while keeping the same measurement interval $\tau$ in Eq.\,\ref{V1} \textcolor{black}{(cf.\, Supplementary Note 2)}. The purified spectral distribution of the bath, $P(\omega)$ in Fig.\,\ref{BathPure}c is thereby changed, which may bring about the QZE to AZE transition of a qubit immersed in this bath: As seen in Fig.\,1c and Fig.\,1d, either the QZE or the AZE regime may be attained on demand by choosing the number of outcomes $n$ in the CT. In particular, the $n = 0$ CT in Eq.\,\ref{V1} yields a single-peak bath spectrum which is conducive to QZE dynamics of the qubit \cite{ref52}, whereas the $n = 4$ CT yields a double-peak spectrum of the bath which tends to AZE dynamics of this qubit \cite{ref40,ref46}. Remarkably, the QZE or AZE regimes of a qubit can be stretched, on demand, over much longer time intervals  than before (Fig.\,\ref{ProbeFID}) because of a high-degree of non-Markovianity of the steady-states of $B$. 

As a signature of the bath purification,  we consider the autocorrelation of qubit measurement  outcomes, $\langle C(t)C(t+t_d)\rangle$, separated by waiting (delay) time $t_d$. We experimentally and theoretically find (see Fig.\,\ref{fig:four}, Methods) that the probability to obtain the same measurement outcomes remains independent of the time $t_d$, i.e. perfect autocorrelation is maintained, until such $t_d$ that are comparable to the bath lifetime, for which the bath gets modified by its own relaxation.

\subsection*{Experiment}
\begin{figure}
\begin{center}
\includegraphics[width=1.0\textwidth]{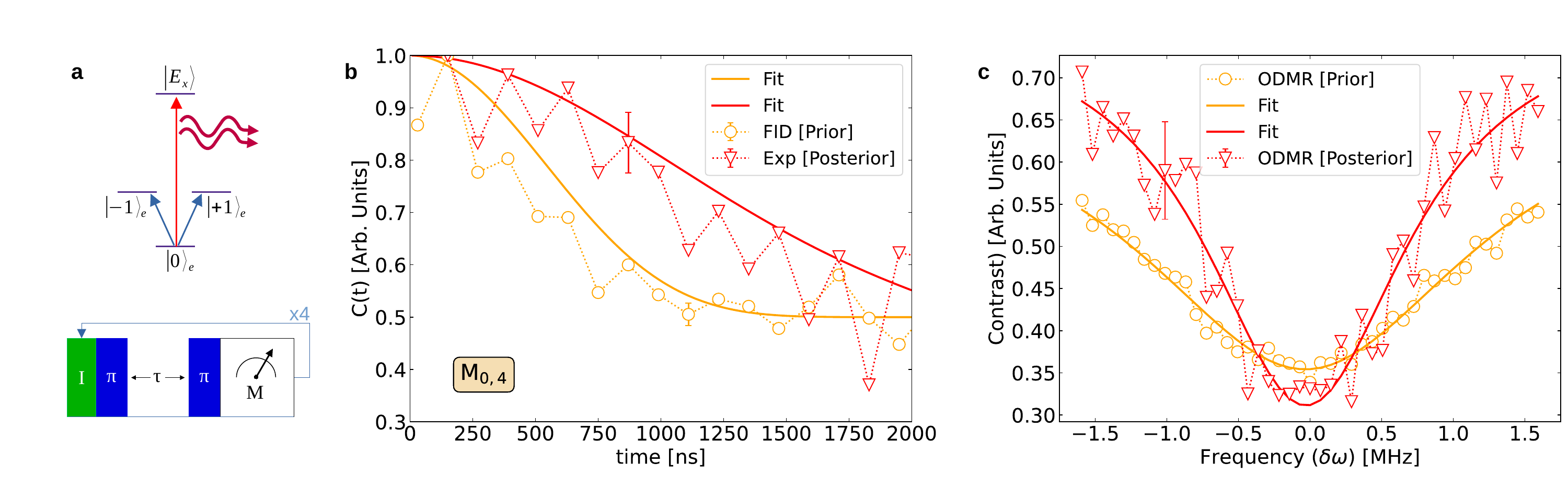}
\end{center}

\vspace{-5mm}
\caption{{\bf Probe FID before and after bath purification:} (a) Energy-level diagram of the central probe-spin (qubit) depicting the basis states \textcolor{black}{($\ket{\pm 1}, \ket{0}$) and the cycling transition between the ground state $\ket{0}$ and the excited state $\ket{E_x}$ used for readout.} Shown below is the pulse-sequence for four measurements. (b) Experimental and theoretical FID of the probe spin before \textcolor{black}{(orange)} and after \textcolor{black}{(red)} four measurements with identical (0) results (CT $M_{0,4}$). The effective FID rate of the corresponding conditional bath state is $4$ times lower than the original \textcolor{black}{(orange)} decay rate indicating the \textcolor{black}{AZE} regime of the qubit. (c) The central ODMR peak of the probe spin reflecting the modification of the initial spin-bath spectrum \textcolor{black}{(orange)} to a narrow single peak and after \textcolor{black}{(red)} the conditional measurement \textcolor{black}{sequence}, CT $M_{0,4}$ in full agreement with the measured and calculated FID in (b). \textcolor{black}{The solid lines are fits to the experimental data and were not obtained from theoretical simulations.} }
\label{ProbeFID}
\end{figure}

In the experiment, the $S$-spin is a low strain NV center in diamond. A small magnetic field is applied along the NV axis [111] to counter the earth’s magnetic field, thereby
eliminating the net $B$-field at the location of the NV center. Hence, the $\ket{\pm 1}$ states in Fig.\,\ref{ProbeFID}a are degenerate. Only the
bright superposition state $\vert + \rangle$ couples to $\ket{0}$, where the two states $\vert \pm \rangle = \frac{[\vert +1\rangle \pm \vert -1 \rangle]}{\sqrt{2}}$ form the two-level probe-spin ($S$) subspace described above. For a projective readout of the probe state, we map the $\ket{+}$ state to the ancillary state $\ket{0}$ via a microwave transition. At a low temperature ($4.2$ K), according to the optical selection rules, we achieve high-fidelity single-shot readout. 

Our protocol starts with the initialization of the electron spin to the state $\ket{0}$ by resonantly exciting it \textcolor{black}{via the $A_1$ transition which causes optical pumping into the $\vert 0\rangle$ state \cite{ref22}},
followed by a microwave transition, $\ket{0}\rightarrow \ket{+}$ \textcolor{black}{(Fig.\,2a)}. We let the prepared state acquire a phase over time $\tau$ due to its interaction with the bath, and then map the population back to the ancillary state $\ket{0}$ and perform single-shot readout using the $E_x$ excitation. This procedure is repeated $m$ times at intervals $\tau$ \textcolor{black}{(here $\tau$ is either $600$ ns or  $1.2\,\upmu$s).}
\begin{figure}
\begin{center}
\includegraphics[width=1.0\textwidth]{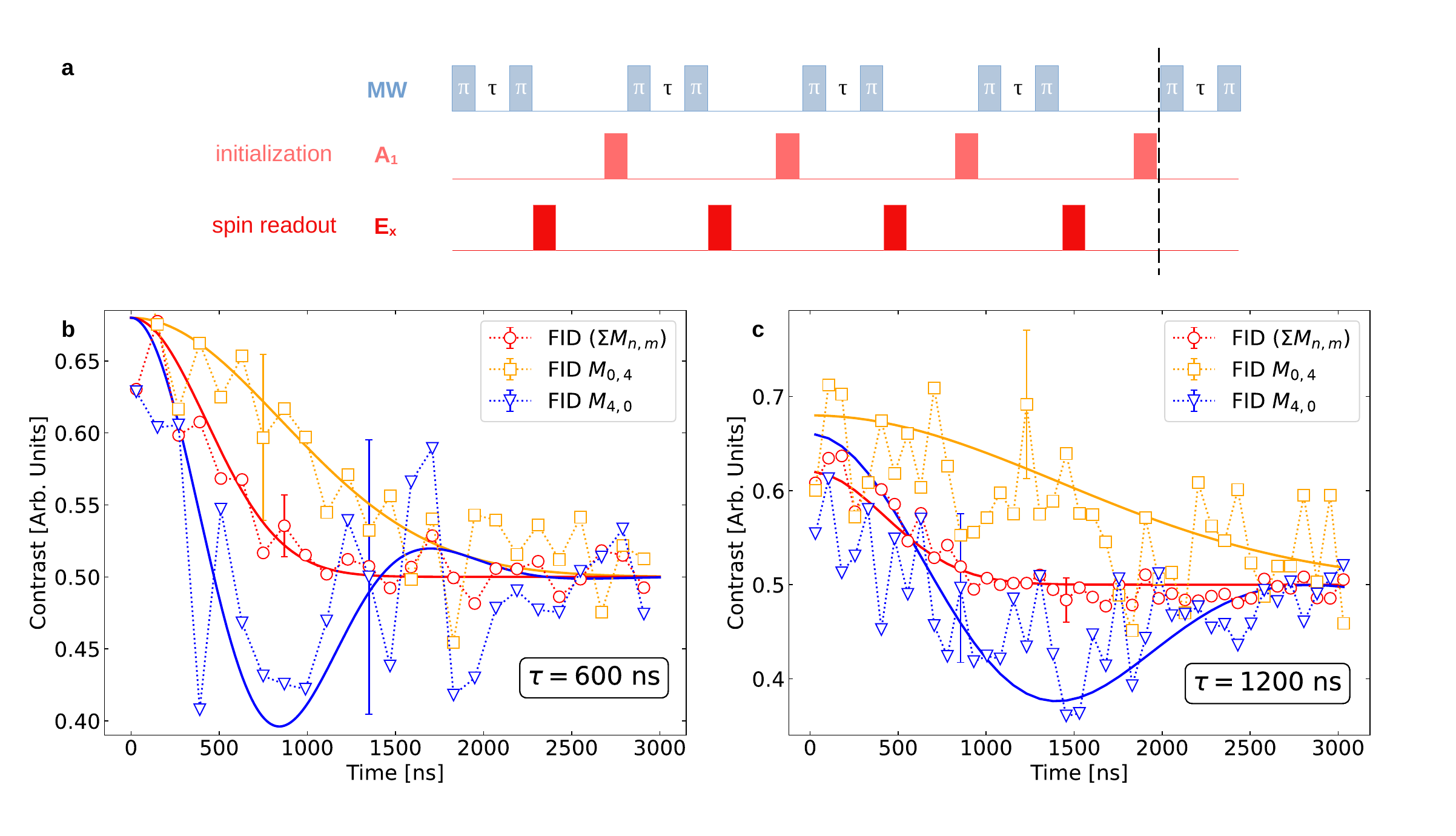}
\end{center}

\vspace{-5mm}
\caption{{\bf Zeno and Anti-Zeno effects on the evolution of the probe-spin following bath purification}(a) The experimental pulse diagram depicting the microwave pulses (blue) interleaved with repeated initialization ($A_1$) and measurements ($E_x$), for preparation of the bath state via CT sequence $M_{n,m}$. Following the $M_{n,m}$ sequence, the FID of the probe spin is measured again and compared with the original FID. (b, c) The FID of the probe spin interacting with the purified bath state by $M_{0,4}$ (orange) and $M_{4,0}$ (blue) is compared with the average FID obtained from all possible conditional bath state preparations (red). The same procedure is repeated for two different measurement intervals $\tau = 600$\,ns (b) and $\tau  = 1200$\,ns. The solid-lines are fits to the experimental data. The experimental plots are in full agreement with the theoretical curves in Fig.\,1c,\,1d. The best fits take the form $C(t) = 1/2 + a e^{-(t/T)^2}\cos(\omega_B t)$, where $a$, $T$ and $\omega_B$ are fitting parameters (for details see Methods). \textcolor{black}{The error bars for the experimental data shown here are few-percent large (see Methods).}} 
\label{fig:three}
\end{figure}

To confirm the actual purification of the bath toward a low-entropy state we perform two different experiments, wherein: 
(i) we measure and compare the probe-spin FID (Eq.\,\ref{Eq:H_in_Sz_basis}) before and after the bath purification by conditional measurements; and (ii) we measure the bath spectrum via the standard procedure of the Optically Detected Magnetic Resonance (ODMR) that records the spin noise spectrum of the probe-spin, before and after the conditional measurements. 

\textcolor{black}{In Fig.\,2b, we see that the FID rate of the probe-spin corresponding to the bath state obtained at the end of the CT $M_{0,4}$ is $4$ times slower than the original decay rate. The corresponding ODMR peak of the probe-spin shown in Fig.\,2c, is narrower than the original broad spectrum, having a well-defined peak. In Fig.\,3 we show the probe FID-s corresponding to the purified bath states obtained after the CT-s $M_{0,4}$ (orange) and $M_{4,0}$ (blue) and compare them with the average FID obtained from all possible conditional bath state preparations (red). The results are shown for two different measurement intervals $\tau = 600$\,ns (Fig.\,3b) and $\tau  = 1200$\,ns (Fig.\,3c).} 

\textcolor{black}{The noise of the probe-spin is affected by the power broadening of the microwave drive, electronic spin noise from P1 centers and the imperfect optical readout, in addition to the nuclear spin-bath. We have not separated the contributions of these noise sources from the noise associated with the nuclear spin-bath. Nevertheless, the results shown in Figs.\,\ref{ProbeFID}, \ref{fig:three} unmistakably reveal that the prolonged coherence time of the probe-spin corresponds to the narrowing down of its spin noise spectrum, after $4$ conditional measurements that purify the bath to a low-entropy state. Namely, sources of noise other than the spin bath are practically unaffected by the conditional projective measurements made on the probe, whereas the nuclear spin-bath, to which the probe is strongly coupled, is drastically changed by the probe measurements. Hence, we can safely attribute the narrowing of the ODMR spectral peak, which is indicative of noise reduction, to the purification of the spin bath. }

The CT success probabilities for the four-measurement case shown in Fig.\,3 are approximately $0.1\%$  \cite{ref71}. They can be strongly increased by $\tau$ optimization \cite{ref77}. The SSRO fidelity for the presented results is on average always $\geq 95\%$. We have chosen here two CTs with the same outcomes but different measurement intervals: one within the correlated (non-Markovian) measurement regime, $\tau < T_2^*$, the other with $\tau = T_2^*$, noting that the bare spin decoherence time $T_2^*$ is relevant in our experiment. For both CTs the observation of a measurement string $M_{0,4}$ projects the bath state to the same mixture of the effective zero-field state ($j=0$) and those high-$j$ states that have survived the purification (Eq.\,\ref{eq:rho_infty}). Since the density of states that correspond to the zero bath-field is exponentially larger, by the factor \textcolor{black}{$^{N} C_{N/2} = \frac{N !}{(N/2)!(N/2)!}$}, than that of the high-$j$ states, the probability of the bath to collapse toward the $j=0$ is high, yielding a single-peak bath spectrum (Fig.\,2) as predicted in Fig.\,1c (orange). The purification is not restricted to particular choices of $\tau$ as long as $\tau \leq T_2$ ($\approx 100 \,\upmu$s), but as $\tau \rightarrow T_1 (\approx 12.9\,$ms) \cite{ref70}, the purification requires an increasing number of measurements, thus making the protocol less practical.

\begin{figure}
\begin{center}
\includegraphics[width=.5\textwidth]{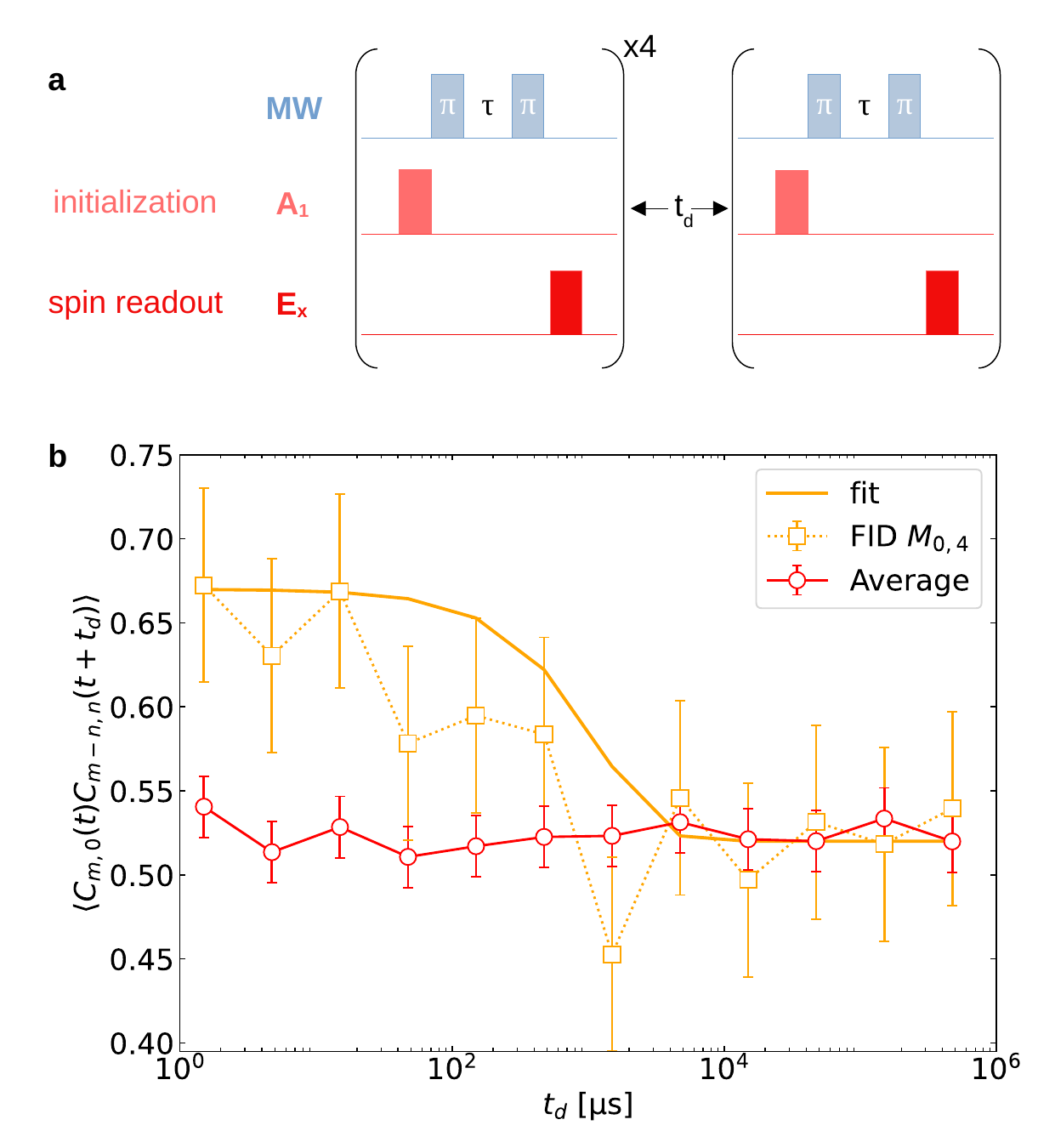}
\end{center}

\vspace{-5mm}
\caption{{\bf Measuring the lifetime of the purified bath-state via qubit-state autocorrelation.}  (a) The pulse sequence used for the  measurement. After the conditional bath state \textcolor{black}{purification by CT} $M_{0,4}$, we allow a waiting time $t_d$ \textcolor{black}{(plotted on log-scale)} and then measure again the probability of obtaining an identical measurement result for the qubit. (b) The probability of obtaining the result $0$ after a long waiting (delay) time $t_d$ for bath state preparations via $M_{0,4}$ (orange)  is shown for measurement intervals $\tau = 1.2\,\upmu$s. The flat part of the orange curve marks a long auto-correlation time for the qubit. These probabilities are compared with the case without bath purification (red), where the auto-correlation time is much shorter. The effective lifetime of the bath, following purification, is seen to extend towards the $T_1^{e}$ lifetime of the probe. Further details on the error bar analysis \textcolor{black}{are} given in Methods. }
\label{fig:four}
\end{figure}
To change the probe dynamics from the Zeno (QZE) to the Anti-Zeno (AZE) regime, we repeat the experiments with a different CT, $M_{4,0}$, and also a different measurement interval $\tau$. In Fig.\,\ref{fig:three}b,\,\ref{fig:three}c we show the FID samplings of the probe-spin following different conditional preparations of the bath by repeated selective measurement samplings of the probe-spin. We see two dramatically different decay profiles of the probe-spin coherence conditioned by the choice of CT: either a slowdown of the decay (red-QZE) or an underdamped oscillation (blue-AZE). In the AZE regime, the probe coherence exhibits underdamped oscillation at the mean frequency of the bath field $\omega_B$,
reflecting the bath preparation in a non-zero $j$-state. \textcolor{black}{Thus, whereas the QZE can be associated with dephasing-time stretching, the AZE corresponds to the preservation of the probe coherence, manifested by its oscillations.}

{\color{black} We note that the data points in Fig.\,3b and 3c are rather noisy mainly because of the very long integration-time limitations set by the occurrence probabilities of the $M_{n,m}$ (see Methods). In Fig.\,4a we show the pulse sequences separated by a waiting time $t_d$ used for measuring the autocorrelation time of a qubit in a purified bath, as discussed in the Theory Section. In Fig.\,4b we plot the measured autocorrelation time as a function of the waiting time $t_d$, obtained after purification via CT $M_{0,4}$ and compare it with the autocorrelation time measured without purification: the autocorrelation \textcolor{black}{persists $\sim 10^3$ times longer after the purification.}}

\section*{Discussion}

We have put forth a quantum control paradigm whereby a thermal bath can be purified to a desired low-entropy state. Such a protocol is demonstrated, theoretically and experimentally, by performing selective measurements of a system (probe spin) coupled to a spin-bath with \textcolor{black}{rms} coupling $\bar{g}$, at intervals $\tau$ compatible with the anti-Zeno (AZE) regime, (\textcolor{black}{in the range $1\geq \bar{g}\tau \geq 0.5$}). As demonstrated here, only at such AZE-compatible (sizeable) intervals can we purify a random, thermally broad bath spectrum, by a sequence of selective measurements of the probe, conditional on its successful projections on one of its basis states (Fig.\,\ref{BathPure} -- \ref{fig:three}). By contrast, intervals $\bar{g}\tau \ll 0.1$, compatible with the quantum Zeno (QZE) regime, merely conserve (freeze) the initial thermal bath spectrum (see Supplementary Note 3). 

We have used this novel protocol to demonstrate, theoretically and experimentally, the ability of an initially fully mixed spin-bath state to attain a steady-state  \textcolor{black}{with either single- or multi-peaked energy-state distribution, by choosing the desired conditional measurement trajectory (CT) of the probe. }

We have revealed remarkable, hitherto unknown, persistent effects of such \textcolor{black}{bath} purification: by preparing an appropriate purified steady state of the bath, we can enforce, on demand, either the QZE or the AZE on a probe-spin qubit immersed in the purified bath. \textcolor{black}{Unlike the existing QZE and AZE protocols \cite{ref2.5, ref48, ref50, ref51, ref52, ref53, ref54, ref55, ref56, ref57, ref58, ref59, ref60, ref61, ref62}, the present protocol provides a handle on the QZE-AZE transition, not only by changing the measurement rates but also by choosing the particular CT. This handle allows us to extend the autocorrelation time of the probe-spin coherence (Fig.\,\ref{fig:four}) by orders of magnitude, determined by the high-fidelity projection of the final state onto the initial state. In Fig.\,4 we see that the bath spectra corresponding to the steady-states can be narrowed down dramatically, which renders any initial state of the probe highly resilient to decoherence. External field strengths and intra-bath interactions can be further tuned to achieve an even broader class of steady states for applications in sensing protocols (cf.\, Supplementary Note 4).}

The demonstrated effects are not observable unless we track a particular
measurement sequence, performed at AZE-compatible intervals, which we dub a conditional trajectory (CT). Otherwise, after following m projective measurements of the spin-probe S, its mean coherence exhibits decay that is an average over the $2^m$ possible measurement outcomes, pertaining to $2^m$ different CT. One can verify \cite{ref70} that the corresponding averaged state of the bath
B is then the same as the unmeasured state at time $t$, i.e., $\sum_N \rho_B(M_N, t) = \rho_B(t)$. Thus, averaging over all possible  free induction decay (FID) patterns resulting from all different CT yields the FID of $S$ in the absence of any measurements. The FID time $T_2^*$, is actually an average of faster and slower decays of the $S$-coherence caused by the bath: There are bath states that almost freeze the S-coherence decay (obeying the QZE) and those that, on the contrary, speed it up (conforming to the AZE). All such effects become observable upon resolving the individual CTs of $S$, as demonstrated here (Figs.\,\ref{BathPure}, \ref{ProbeFID}, \ref{fig:three}). These CTs correspond to specific states of $B$ and hence to particular magnetic (dipolar) field distributions generated by the bath. 

\textcolor{black}{Remarkably, we have shown that a desired steady state of the bath is attained with appreciable success probability that is nearly independent of the number of measurements $m$, in sharp contrast to the standard KLM protocol \cite{klm}, due to the correlation of successful outcomes of the CT as the B-distribution narrows down [Fig.\,\ref{BathPure}(b)]. This improved scaling of spin-bath purification probability with the number of successful outcomes can be further improved by a protocol that is akin to optimized conditional measurements introduced by us to control the evolution of quantum observables \cite{ref70,ref71,ref77}.}

A spin bath purified by our method can be advantageous for quantum sensing and quantum computation: unlike existing dynamical control of qubits, which is in effect only during the pulse sequence, such as dynamical decoupling (DD)  \cite{ref35,ref37,ref38,ref39}, or the bath-optimized minimal-energy control (BOMEC) devised by us \cite{ref42}), our measurement-induced steady-state of the bath persists long after the control has ceased and therefore allows us to \textcolor{black}{perform coherent operations on} other qubits immersed in the bath over considerably long time scales (Fig.\,\ref{fig:four}). 

An important application to quantum sensing is foreseen for the \textcolor{black}{CTs} that yield pronounced coherence oscillations of the probe spin in the AZE regime (Fig.\,\ref{fig:three} and Supplementary Note 4 \textcolor{black}{ }): such oscillations allow us to identify  bath spins that strongly-interact with the probe via their high $j$ (high-field) state. Such states are currently identified through complex correlation spectroscopy methods \cite{ref71, ref72}. \textcolor{black}{For example, from the underdamped AZE oscillations in Fig.\,\ref{fig:three}, we identify the presence of a nuclear spin-bath state whose large coupling strength with the NV probe is  $\sim 500 \mathrm{kHz}$. Since the individual couplings to each nuclear spin cannot be resolved, we only obtain the collective effects of the bath.}

To conclude, we have introduced handles on spin-bath purification as a method of coherence preservation of qubits in such baths beyond the limitations of existing dynamical control that leaves the bath unchanged \cite{ref35,ref37,ref38,ref64,ref66,ref67}. These handles are also prerequisites for the exploitation of spin baths prepared in purified states whose lifetime can be extended by few orders of magnitude, as a quantum memory. Their
reliable long-term storage of quantum information  \cite{ref70,ref71,ref72,ref73} is \textcolor{black}{a} functionality compatible  with  quantum  network operations \cite{hansen-r74,ref74a}. Alternatively, by virtue of their low entropy, such purification of spin baths \textcolor{black}{is} expected to substantially  increase  the  signal-to-noise  ratio  of  quantum  sensing \cite{ref10,ref13,ref14,ref15,ref16,ref17,ref18,ref19,ref20,ref21,ref22,ref23,ref24,ref24.10,ref25,ref26,ref27,ref28,ref29,ref32}.

 \section*{Methods}

To interpret the autocorrelation, let us consider a  bath state purified at $t = m\tau$ by the CT $M_{n,(m-n)}$ and evaluate the probability for the modification of this bath state by relaxation modeled by a depolarizing channel \cite{ref11}. The bath state at later time $t + t_d$ is then given by
\begin{equation}
   \rho_B^{\lbrace n,(m-n)\rbrace}(t+t_d)=[1-p_B(t_d)]\rho_B^{\lbrace n,(m-n)\rbrace}(m\tau)+p_B(t_d)~\mathbf{\hat{I}}/2^N.
\end{equation}
where $p_B(t_d) = 1 - \exp{(-\Gamma t_d)}$, and $\Gamma$ is the relaxation rate of the bath. \textcolor{black}{This formula assumes that the bath relaxation is Markovian over the time interval $t_d$.} 

With this bath state, one can evaluate the probability of selecting a qubit in states $\ket{\pm}$ at $t_d$ beyond the bath correlation time, as discussed below. Following the initialization of the qubit to the $\ket{+}$ state after the $m^{\rm th}$  measurement, the full ($S$+$B$) density matrix $\rho$ at time $t + t_d$ is given by

\begin{equation}
    \rho(t + t_d) = \hat{U}_{S + B}(t_d)\,\Big[ \vert + \rangle\langle + \vert \otimes \rho_B^{\lbrace n,(m-n)\rbrace}(t) \Big] \, \hat{U}_{S + B}^{\dagger}(t_d).
\label{eq:rho_m_tau_plus_t}
\end{equation}
Upon tracing out the bath, we can infer from Eq.\,\ref{eq:rho_m_tau_plus_t} the autocorrelation of qubit state  measurements separated by $t_d$. For $\Gamma t_d \gg 1$ the autocorrelation is determined  by the probabilities of the $\ket{\pm}$ states

\begin{equation}
    p_{\pm}(t + t_d) \approx \frac{1}{2^{N + 1}}\;\sum_j \Big[ 1 \pm \cos(2\omega_j t_d) \Big].
\end{equation}
which  are in full agreement with the experimental results in Fig.\,\ref{fig:four}. They show that the qubit coherence autocorrelation persists after the bath purification ca.\,1000 times longer than in the absence of such purification \textcolor{black}{and is limited} by  the nuclear-spin \textcolor{black}{lifetime in the bath ($T_1 > 10\,\mathrm{s}$).} 

In Fig.\,\ref{fig:four} long time delay $t_d$ is introduced between the bath purification by conditional measurements and the subsequent samplings of the probe-spin qubit coherence. The measurement of the probe-spin coherence is almost unchanged up to a millisecond and then slowly decays towards the fully mixed state. This corresponds to a very long lifetime of the purified bath spin-state.

Remarkably, we observe a persistent autocorrelation of the probe coherence over very long time delay $t_d$, Fig.\,\ref{fig:four}(b), which is dependent only on the CT, $M_{0,4}$ or $M_{4,0}$, and not on the sampling timescale: thus, whereas the original QZE and AZE depend on time-resolution, and are caused by non-selective independent measurements, here we observe the opposite: their autocorrelation counterparts depend on the energy resolution of the bath field and are caused by state-selective measurements. 
 
{\color{black}
 \subsection*{Experimental error bars}

The measurements, performed at low temperatures, were done using a continuous-flow liquid helium cryostat, which entails a periodic Dewar replacement. This limits the measurement time to around $45-50$ hours, resulting in insufficient measurement statistics. These statistical error bars exceed $>5\%$ and dominate all other errors in the experiment. 

The average FID curve in Fig.\,3b has a low statistical error, while the $M_{0,4}$ and $M_{4,0}$ CTs are only averaged over $300$ and $110$ counts, respectively, leading to error bars of $\sim 6- 10\%$. In Fig.\,3c, the average FID curve has an error bar of $2.4\%$, while the CTs $M_{0,4}$ and $M_{4,0}$, both have the same error bar ($8\%$). This can be attributed to the complete decoherence of the probe for the measurement interval $\tau = 1200$\,ns, leading to identically probable occurrence of both $'0'$ and $'1'$ measurements. In contrast, for $\tau = 600$\,ns, in Fig.\,3b, the occurrence probability of $'0'$  measurement results is higher than that of $'1'$ results. 

In Fig.\,4, the error bar for the correlation curve for the CT $M_{0,4}$ is $\sim 6\%$, as observed for the $300$ coincidence (correlation) counts shown here. 

Since the success probability of the CT $M_{0,4}$ is higher than that of the $M_{4,0}$, the corresponding error bar is smaller. With the advent of cryogen-free cooling systems that can continuously operate over many weeks, we envisage that in future experiments the statistical errors will be suppressed.}

 \subsection*{Fitting Parameters}
Below we summarize the curve fitting parameters for the experimental data shown in Fig. 3 and Fig. 4. 
 \\
{\bf Figure 3a}: All solid-line fits to all the curves take a simple analytical form $C(t) = 1/2 + a e^{-(t/T)^2)}\cos \omega_B t$, with $a$, $T$ and $\omega_B$ being the fitting parameters. For the $M_{4,0}$ sequence $a =0.18 $, $T =1.2$ ns and $\omega_B=530$\,kHz. Similarly for $M_{0,4}$ sequence $a =0.18 $, $T =1.2\,\upmu$s and $\omega_B=0$. For the average FID curve $a =0.18 $, $T =0.6\,\upmu$s and $\omega_B=0$.
 \\
{\bf Figure 3b}: With a similar analytical fit function given above we find (i) for the $M_{4,0}$ sequence $a =0.16 $, $T =2.37\,\upmu$s and $\omega_B=313.5$\,kHz, (ii) for $M_{0,4}$ sequence $a =0.18$, $T \sim~ 2$\,ns and $\omega_B=0$ and (iii) for the average FID curve $a =0.12 $, $T =0.6$\,ns and $\omega_B=0$.
\\
{\bf Figure 4}: The solid-line fit (orange line) takes the exponential decay form $1/2+a\exp(-bt)$ with $a,~b$ as fitting parameters. The best fit for the experimental data are $a = 0.1448$ and {$1/b = 0.2826$\,ms.}

{\color{black}

 \section*{Data Availability}
Source data for all the figures are provided with this paper. Any other data supporting the findings of this paper are available from the corresponding author upon request.
}

\section*{Figure Legends}

\begin{enumerate}
    \item {\bf Bath purification by conditional selective measurements and its posterior (persistent) effects} (a) Schematic description: Conditional selective measurements of the state of the central probe-spin (S) \textcolor{black}{surrounded by a spin bath in the star configuration. The conditional measurements consist of photon-emission detection from the decayed state or non-detection from the initial state. A sequence of measurement events} collapses the spin-bath toward a low entropy state with resolvable, partly-\textcolor{black}{(red)} or \textcolor{black}{fully-polarized} (orange) spins. (b) The success probability, $p_m$ of $m$ positive measurement outcomes: solid red -- for a probe subject to uncorrelated (Markovian) measurement sequence (solid red) that leave the spin-bath unchanged, dashed orange -- for a probe subject to  correlated measurements that purify the bath. The success probability exponentially decays with $m$ for uncorrelated measurements, while it saturates for correlated measurements. (c) Simulated bath spectrum $P(\omega_j)$ plotted for the cases prior to measurements (red -- thermal bath) and after the conditional (selective) measurement sequences (CT) $M_{0,4}$ (orange) and $M_{4,0}$ (blue).  Both (orange and blue) posterior spectra are much narrower than the prior spectrum, indicating low-entropy steady-states of the bath. The orange spectrum is peaked at bath states with zero magnetic field, whereas the blue spectrum is peaked at higher magnetic field values of the bath spins. (d) Probe coherence decay  prior (red) and posterior to measurements (orange, blue) for the spectra shown in (c). The orange curve is in the quantum Zeno (QZE) regime (decay slowdown) while the blue curve is in the anti-Zeno (AZE) regime (decay speedup). These curves are simulated for a bath of $N=8$ spins that are randomly (inhomogeneously) coupled to the probe spin. The timescale between measurements is chosen to be $\tau = 0.6\,\upmu$s, such that it is close to the bare $T_2^*$ obtained from the prior FID (red solid-line in 1(d)). (See Results, SI)
    
    \item {\bf Probe FID before and after bath purification:} (a) Energy-level diagram of the central probe-spin (qubit) depicting the basis states \textcolor{black}{($\ket{\pm 1}, \ket{0}$) and the cycling transition between the ground state $\ket{0}$ and the excited state $\ket{E_x}$ used for readout.} Shown below is the pulse-sequence for four measurements. (b) Experimental and theoretical FID of the probe spin before \textcolor{black}{(orange)} and after \textcolor{black}{(red)} four measurements with identical (0) results (CT $M_{0,4}$). The effective FID rate of the corresponding conditional bath state is $4$ times lower than the original \textcolor{black}{(orange)} decay rate indicating the \textcolor{black}{AZE} regime of the qubit. (c) The central ODMR peak of the probe spin reflecting the modification of the initial spin-bath spectrum \textcolor{black}{(orange)} to a narrow single peak and after \textcolor{black}{(red)} the conditional measurement \textcolor{black}{sequence}, CT $M_{0,4}$ in full agreement with the measured and calculated FID in (b). \textcolor{black}{The solid lines are fits to the experimental data and were not obtained from theoretical simulations.}
    
    \item {\bf Zeno and Anti-Zeno effects on the evolution of the probe-spin following bath purification}(a) The experimental pulse diagram depicting the microwave pulses (blue) interleaved with repeated initialization ($A_1$) and measurements ($E_x$), for preparation of the bath state via CT sequence $M_{n,m}$. Following the $M_{n,m}$ sequence, the FID of the probe spin is measured again and compared with the original FID. (b, c) The FID of the probe spin interacting with the purified bath state by $M_{0,4}$ (orange) and $M_{4,0}$ (blue) is compared with the average FID obtained from all possible conditional bath state preparations (red). The same procedure is repeated for two different measurement intervals $\tau = 600$ ns (b) and $\tau  = 1200$ ns. The solid-lines are fits to the experimental data. The experimental plots are in full agreement with the theoretical curves in Fig.\,1c, 1d. The best fits take the form $C(t) = 1/2 + a e^{-(t/T)^2}\cos(\omega_B t)$, where $a$ is a fitting parameter. The bath-field \textcolor{black}{strength} obtained from such fitting is \textcolor{black}{either} $\omega_B=200$ kHz \textcolor{black}{or} $\omega_B = 80$ kHz, and $a \approx 1$. \textcolor{black}{The error bars for the experimental data shown here are few-percent large (see Methods).}
    
    \item {\bf Measuring the lifetime of the purified bath-state via qubit-state autocorrelation.}  (a) The pulse sequence used for the  measurement. After the conditional bath state \textcolor{black}{purification by CT} $M_{0,4}$, we allow a waiting time $t_d$ \textcolor{blue}{(plotted on log-scale)} and then measure again the probability of obtaining an identical measurement result for the qubit. (b) The probability of obtaining the result $0$ after a long waiting (delay) time $t_d$ for bath state preparations via $M_{0,4}$ (orange)  is shown for measurement intervals $\tau = 1.2\,\upmu$s. The flat part of the orange curve marks a long auto-correlation time for the qubit. These probabilities are compared with the case without bath purification (red), where the auto-correlation time is much shorter. The effective lifetime of the bath, following purification, is seen to extend towards the $T_1^{e}$ lifetime of the probe. Further details on the error bar analysis \textcolor{black}{are} given in Methods. 
\end{enumerate}

\begin{acknowledgements}
\noindent D.\,D.\,B.\,Rao, G.K. and J.\,W.\,acknowledge the support of DFG (FOR2724), ERC project SMeL (742610), DFG SFB/TR21, EU ASTERIQS (820394), QIA (820445), Max Planck Society, the Volkswagenstiftung, the Baden-W\"urttemberg Foundation.\\ S.\,Y.\,acknowledges the financial support from Hong Kong RGC (GRF/14304618). A.\,F.\,is the incumbent of the Elaine Blond Career Development Chair and would like to acknowledge the historic generosity of the Harold Perlman Family, research grants from the Abramson Family Center for Young Scientists and the Willner Family Leadership Institute for the Weizmann Institute of Science, as well as support from the Israel Science Foundation (ISF 963/19).
\end{acknowledgements}

\section*{Author Contributions}
SY, DDB and JW designed the experiments. SY performed the experiments.
DDB, GK and AC conceived the theoretical model and performed the calculations. SY, DDB and AF analyzed the experimental data. JW, GK and DDB supervised the entire project. GK, DDB and AC wrote the manuscript with input from all authors.

\vspace{2cm}

\section*{Competing interests}

We declare no competing interests.

\vspace{2cm}

\end{document}